\documentstyle[11pt]{article}

\begin{document}

\title{CoRR: A Computing Research Repository%
\thanks{This article is based on (and borrows
liberally from) two earlier articles:
``A Computing
Research Repository'', D-Lib Magazine, November 1998
(http://www.dlib.org/dlib/november98/11halpern.html) and
``The Computing Research Repository: Promoting the Rapid Dissemination
of  Computer Science Research'',
{\em Proceedings of ACM Digital Libraries '99}, 1999, pp.~3--11.  The
latter article is joint with Carl Lagoze; I thank him for all his
contributions.
This work was supported in part by the NSF, under grant
IRI-96-25901.}}
\author{Joseph Y. Halpern\\
Computer Science Department\\
Cornell University\\
halpern@.cornell.edu}
\date{\today}
\maketitle

\section{Introduction}
Computing research relies heavily on the rapid dissemination of results.
As a result, the formal process of submitting papers to journals has
been augmented by other, more rapid, dissemination methods.  Originally
these involved printed documents, such as technical reports and
conference papers.  With the advent of the Internet, researchers
developed and began to use a variety of electronic means for rapid
dissemination.  Individual and organizational web sites made it possible
to provide cheap and almost instantaneous access to research results.
But these resources were fragmented.  There
was no single repository to which researchers from the whole field
of computing could submit reports, no single place to search for
research results, and no guarantee that information would be archived
at the end of a research project.

This changed in September 1998.  Through a partnership of ACM
(http://www.acm.org), the  LANL (Los Alamos National Lab) e-Print archive
(http://xxx.lanl.gov), and NCSTRL (Networked Computer Science Technical
Reference Library -- http://www.ncstrl.org), an online Computing Research
Repository (CoRR) was established.  The Repository is available to all
members of the community at no charge.  They can submit papers, browse
and search papers currently on the Repository, and subscribe to get
notification of new submissions.

In the rest of this article, I briefly describe how CoRR was
set up and discuss some policy issues.

\section{Setting up CoRR: Issues and Decisions}

ACM was (and continues to be) interested in experimenting with different
approaches to disseminating research.
In May 1997, a committee was formed under
the auspices of the ACM Publications Board to consider one
such change: setting up an online repository for computing research.%
\footnote{
Appendix A gives the membership of the committee, which consisted
mainly of people active in digital libraries and electronic publishing.}
Initially, the main focus of the committee's discussions revolved around
the design of the architecture.  Three main options emerged.

The first option was to become part of the LANL repository.
LANL started as a repository for high-energy physics eprints in 1991, several
years before the introduction of the Web.  It pioneered the concept of an
open-access repository for fast publication of scientific research. By
eliminating the time consuming and expensive process of peer review, it has
transformed the dissemination of research in several disciplines.  It
now covers most of physics and has expanded to include repositories for
nonlinear sciences, mathematics, and computation and language.  The LANL
archives are sometimes called a ``pre-print'' service, and indeed many
of the
eprints are subsequently published in conventional journals, but they
are intended as long-term archives, with much greater permanence than
typical Web sites.

As a base for a computing repository, LANL has many attractive
features. Perhaps the most important is that it
clearly works and works well.  It now has
over 100,000 eprints, is growing at the rate of about 25,000/year,
handles
over 100,000 transactions/day, and has over 35,000 users.  Thanks
to funding from the Department of Energy and the National Science
Foundation, it also has a full-time staff.  It is
mirrored in 15 countries, has reasonable search facilities, and offers
services such as email notification of new submissions of interest.

The ACM committee decided against this option primarily because the LANL
interface was not open, in the sense that it did not provide an
interface to which other repositories could link and was not easily
amenable to extension and enhancement.

The second option considered was to become a node in NCSTRL.
NCSTRL is essentially a
common interface to the technical report collections of its (currently
over 100) member institutions.  It has been funded by DARPA and the
National Science Foundations, with most of the technical work recently
being carried out at Cornell University.  The most important features of
NCSTRL from our point of view were that it was explicitly designed with
an open interface and it was a computer science effort.  On the other
hand, NCSTRL did not have all the software necessary for running a
repository.

The third option was to build a new system from
scratch.  This had the obvious advantage that we could design our
own system, which hopefully would have exactly the attributes
we required, but had the equally obvious disadvantage that it would take
time, money, and expertise.

The committee
settled on a hybrid approach that combines the best features of LANL
and NCSTRL, and secured the cooperation of the two groups.
This allowed us to use the well-tested LANL software for submission,
notification, and searching, while still taking advantage of the NCSTRL
architecture.  The NCSTRL architecture makes it
easy to build new gateways from which to access the files, with a
more user-friendly interface and new features.  From the point of
view of the NCSTRL interface, LANL is now just a node on NCSTRL.

We anticipated then that our use of an open protocol
would encourage other scholarly archives to join in this framework,
resulting in a global multi-disciplinary research collection that
could have substantial impact on the nature of scholarly publishing.
Some recent developments have made that hope more likely to become
reality; see Sections~\ref{sec2} and~\ref{sec3}.

With the major decision out of the way, there were
still a number of other important decisions that had to be made
regarding how CoRR would operate:

\begin{itemize}
\item {\em How should CoRR be organized?}  The physics and mathematics
archives at LANL are organized into a relatively small number of subject
areas---38 in the case of physics and 31 in the case of mathematics.
These subject areas play a number of roles.  From the perspective of
document submission, they form the basis for moderation; that is, for
each subject area, there is a moderator who checks submitted papers for
relatedness to the subject (although not quality or novelty).  At the
user-interface level, the subject classes are used as aids for
searching, browsing, and subscribing.

The committee had to decide how to partition the computing field into
subject areas.  One choice was to use the ACM classification system (see
http://www.acm.org/class/1998/overview.html).
The ACM classification scheme has the advantage of being a
relatively stable scheme that covers all research in computing, which
has been carefully crafted over the years.  Unfortunately, it doe not
map too well to the current major areas of academic computer science.
In particular, it seemed difficult to find moderators for subject areas
that corresponded to major topic under the ACM classification system.

In the end, the committee chose to use both approaches.  Authors are
asked to classify papers both by
by choosing a subject area from a list of subject areas (of which there
are currently 33; see http://xxx.lanl.gov/archive/cs/subj.html for a
description of the areas) and by choosing a primary classification from
among the roughly 100 third-level headings in the 1998 ACM Computing
Classification System (see http://www.acm.org/class/1998/overview.html).
While the
subject areas are not mutually exclusive, nor do they (yet) provide
complete coverage of the field, they seem to
better reflect the active areas of research in CS.  Each subject area
has a moderator.  While documents are partitioned by subject area,
readers can search and get notified of new papers both by ACM
classification and subject area.
We expect to add more subject areas, subdivide current subject areas,
and perhaps delete subject areas, depending on demand over time.
(Interestingly, the committee that formed the mathematics archive at
LANL
chose not to use the AMS classification scheme at all, instead opting
for their own subject classification; see
http://xxx.lanl.gov/new/math.html.)

Notice it is authors who choose subject areas.
By using the LANL software, we were effectively committing to the LANL
approach for paper submission.  The LANL philosophy is to have
authors do the submission, with as much automated as possible.
Authors send their documents to the LANL repository, by email, by ftp,
or by using Web interface provided by LANL. They are expected to
provide their paper in specified formats, provide
an abstract, and to classify their papers by subject area and ACM class.

\item {\em What about copyright?}
Publishers typically require authors to transfer copyright when they
publish a paper, since it gives them more freedom of action and more
control of the disposition of the paper.  The committee decided not to
require any transfer of copyright or publication rights.
Authors continue to retain copyright when they submit (although
they may have to transfer
rights if they wish to publish in certain journals).

\item {\em How long should papers stay on CoRR?}
The committee viewed
CoRR as archival; the expectation is that papers
submitted will stay there permanently.  This does not prevent authors
from updating their papers.  Updated versions of a paper
can be posted at any time, but versions not removed or changed within
24 hours of submission will remain on the repository as well.  All
versions are timestamped, to avoid confusion.  The most recent version
of a paper is the one accessed by default, but there are
pointers to the earlier versions.  This prevents a situation where, for
example, author A improves on the results of an early version of B's
paper, but finds that these improvements seem foolish when the only
version of B's paper that is available has better results.

\item {\em What submission formats should be accepted?}
For many years, physicists have used TeX as the standard format for
research papers, because of the control that it provides for representing
mathematics.  Therefore, the LANL archives provide excellent support for
several versions of TeX.  Theoretical computer scientists also use TeX,
while PostScript has been a favorite format for computing technical
reports.  Currently, authors can submit documents to CoRR using
Tex/LaTeX/AMSTeX, HTML+GIF, PDF, or Postscript.  However, if TeX
(or one of its variants) is available, it is strongly preferred to
Postscript or PDF (see http://xxx.lanl.gov/help/faq/whytex for the
reasons).  If an author has generated Postscript
or PDF from some variant of Tex, it will be rejected in favor of the Tex
source. (This policy has generated some controversy; see
Section~\ref{sec2}.)

\item {\em What about preservation?}
Long-term preservation of documents in CoRR is clearly a serious
concern.  There are two orthogonal aspects to this issue.  One is the
concern that disk crashes and other software and hardware failures will
cause the loss of many documents; the other involves changing platforms.
The committee felt that LANL's size would prove to be a significant
advantage in dealing with both concerns.  Since LANL has 15 mirror
sites and performs frequent backups
there should be more than enough redundancy to deal with software and
hardware problems.  No one has a definitive answer to the problem of
platform changes, but with so many papers on LANL already, our hope is
that there will be enough pressure that, whenever platforms change,
software will be written to automatically convert the
files in CoRR to whatever platform is current, just as there
is now software to convert Postscript to PDF.

\item {\em Who should participate in CoRR?}
Anyone can browse CoRR, search for papers, and download papers.  The
LANL software permits anyone from a university,
government research lab, or industrial research lab to submit
papers.  (It assumes that anyone coming from a~.edu domain is from a
university; the software recognizes the domain names of the major
government and industrial research labs.)  Others may also submit upon
request, with some minimal argument that they are engaged in research.
This filter was put in place to make it easier to reject ``crackpot''
physics papers.  It made sense when the LANL archive consisted only of
physics papers, especially because it is relatively easy to
characterize the places where research in physics is carried out
This is not quite as easy for computer science.
CoRR is certainly not intended to be
exclusive and we may have to revisit this issue at some point, but so
far it seems not to have been a problem.
There have been very few requests to submit papers to CoRR that have
come from non-recognized domains and no one who has wanted to submit to
CoRR has been prevented from doing so.

\end{itemize}

\section{Issues Raised Through Experience}\label{sec2}

CoRR has been in operation since Sept. 15, 1998.  After an initial
flurry of over submissions, the recent rate has been slightly under one
submission a day. Currently there are about 1800 papers on CoRR, with
about 900 coming from a previously-existing archive at LANL on
Computation and Language, which has now been folded into CoRR. These are
combined through the NCSTRL interface with over 27,000 other
papers. It is too early to tell whether CoRR will really catch on.
However, a number of groups have tentative plans to migrate
archives to CoRR and to use CoRR as a repository for journal and
conference activities.
The prognosis seems good.

A number of issues have arisen in the first year of operation which I
briefly discuss here.

\begin{itemize}
\item {\em Insistence on Source Files:}
There have been bitter complaints
from users about the insistence on TeX source. There seem to be
two orthogonal reasons for these complaints.
The first involves ease of submission. There is no question that it
much easier to submit a single Postscript file than it is to submit a
TeX file and a number of auxiliary files, such as figures in
postscript, a bibliography file, and macro files.  (Note that LANL
automatically rejects TeX files that do not include all the necessary
auxiliary files, pointing out which files are missing.)  A script
has recently been posted on CoRR that automatically collects all the
relevant files, sparing authors the burden of doing so, thus mitigating
this concern.
The second reason is that authors are concerned with the fact that
source files are available for download (although, in fact, they are
rarely downloaded---readers far prefer Postscript or PDF). Authors
are concerned that the availability of the source will make
plagiarism easier and also give readers access to comments that were
intended to be private.  There is a script available at CoRR that will
strip comments from files.  Authors also
now have the option of
making their source unavailable to readers, although CoRR
will still retain the source so as to be able to use it
to convert to new formats as they arise.
The advisory committee
encourages authors to make the source freely available;
one never knows
what other uses for the source will be found in the future.
In any case, it is certainly possible to plagiarize
even from PostScript, given the availability of postscript to ASCII
converters.

It is interesting to note that the physics community (which
presumably has much the same concerns as the computing
community) has been submitting TeX source files to LANL for years. This
suggests that there may be cultural differences between the
communities; perhaps in time the computing
community will also become more comfortable with submitting
source files.

\item {\em User Interface:}
There have been many (legitimate!) complaints about the user interface
at CoRR (and, more generally, at LANL).  Anecdotal evidence suggests
that submitting a paper for the first time can easily take over 45
minutes and be a very frustrating experience.  Things certainly go
much faster and more smoothly with experience, and the new tools
introduced
for TeX submissions should help, but there is no question that the user
interface could stand improvement.  However, designing a good user
interface takes time and energy and CoRR is a volunteer effort.
I hope and expect that over time a better interface will
be developed (or interfaces---the open architecture makes it
straightforward to access CoRR through various gateways, each of which
could have their own interface).

\item {\em Funding:}
Currently, CoRR is riding on the coattails of NSF and
DARPA funding provided to LANL and NCSTRL, and this should
suffice for the foreseeable future. The long-run funding situation
is not yet clear.  In any case,
providing the basic repository services does not seem to be an
expensive proposition.  Of course, new development can be expensive,
but we should be able to take advantage of work done by other
projects, so it may not be not be necessary to do too much development
in-house.  Clearly, when a resource becomes as important to
a community as the LANL archives are to physicists (and I hope
that CoRR will be to computer scientists), that community will
collectively work to ensure funding. However, the economic
models for electronic scholarly publishing (and for the Internet as
a whole) are still the subject of considerable investigation.
It is not clear how the funding issue will be resolved in the
long run.

\item {\em Journal publication:}
There are fields (such as
medicine and chemistry) for which publishers will not publish
papers that have appeared on the web (even on an author's
personal web site).  This seems to be easier to do in cases where one or
two journals dominate a field.  It is harder in fields where there are
many smaller journals, without one dominant one.  In particular, it
has not been the case in
computer science, and is unlikely to become so. Researchers have come to
expect that they will be able to make their papers available rapidly
at online sites, such as CoRR, while still submitting their papers to
conventional journals. Publishers may insist, as part of
their copyright policy, that a paper be withdrawn as a precondition
to journal publication.  In this case, at the author's request, the
paper will indeed be withdrawn.  However, this does not seem likely to
be a problem in Computer Science.  All the major publishers I have
checked with so far (ACM, IEEE, SIAM, Elsevier, Academic Press, and
Springer) allow authors to post
post final versions of their papers on the personal web pages and
preprints on public repositories, such as CoRR, and do not require
preprints to be withdrawn.  Society publishers (ACM, IEEE, and SIAM)
also allow authors to post final (journal) versions of their papers on
CoRR, with proper copyright notice.%
\footnote{CoRR allows authors to indicate where
papers have been published and add links to publisher's digital
libraries.}
(In the case of ACM, this is a two-year experiment;
ACM is examining the impact that this will have on journal sales.
I hope and expect that the policy will be continued after the two-year
experimental period.)
As a result of negotiations with the editorial board
of {\em Artificial Intelligence}, Elsevier is allowing final versions of
papers that appear in that journal to appear on CoRR.  (Elsevier
apparently does not allow this for other journals.)
Even in the biological and medical sciences, there are
indications that things may change, with the advent of PubMed Central
(see
http://www.nih.gov/welcome/director/pubmedcentral/pubmedcentral.htm).
\end{itemize}

\section{Where do we go from here?}\label{sec3}
This is a period of change in scholarly publishing and nobody can
predict the changes that will happen over the next few years. The
impact of CoRR and similar efforts on conventional journals is,
no doubt, a question that many journal publishers are asking.
There are a number of possibilities. In one scenario eprint
repositories such as CoRR could coexist with the conventional
journal model (recognizing that that model will undoubtedly move
to electronic dissemination). For example, LANL has been
providing eprint archives in physics since 1991 without apparent
impact on conventional journals. In another scenario, efforts such
as CoRR could be the primary repository for journal papers, while
providing the foundation for a new and enhanced
role for conventional publishers.

A bare-bones journal could be built that would simply be
a collection of pointers to documents in CoRR and other federated
repositories. The journal would have an editorial board just as
journals do now. Papers would be peer reviewed in the usual way.
Rather than (or in addition to) coming out in print, once a paper
was accepted, the final version would be deposited in CoRR, and
there would be a pointer to it from the journal's web site. There
currently is one such overlay journal in physics: {\em  Advances in
Theoretical and Mathematical Physics}
(http://www.intlpress.com/journals/AMTP). In Computer Science, the
{\em Journal of Artificial Intelligence Research\/}
(http://www.cs.washington.edu/research/jair) is also planning a similar
move.  Publishers could provide value
added services to authors and readers such as summarization
services, advanced searching tools, awareness, and filtering
services that build on the content in CoRR-like repositories.
Clearly, if efforts such as CoRR develop into the primary vehicle
for dissemination of research results, they could significantly
change the business model for scholarly publication.

The structure of CoRR makes it possible to experiment with
other ways of filtering papers besides traditional peer review.
One approach that has often been suggested is that of having a
comment facility. Certainly a comment facility could be added
(either in CoRR itself or in an overlay site). Such
comment facilities have met with mixed success in the past. For
example, there has been little usage of the comment facility
provided by the {\em Journal of Artificial Intelligence Research}; the
comment facility at the {\em Electronic Transactions on Artificial
Intelligence\/} (http://www.ida.liu.se/ext/etai)  has seen more usage.
It remains to be seen whether they will really catch on.

Yet other forms of filtering, such as recommendations from
respected members of the community, are also possible. Indeed,
one such experiment, the {\em ACM SIGMOD Digital Review}, is being
planned now, and will be an overlay over CoRR.

There is clearly a need for value-added services besides
filtering. As it stands, the only attempt to organize the
papers in CoRR is by means of subject classification and ACM
classification. These classifications are
provided by authors, without the involvement of librarians.
Getting librarians
actively involved in cataloging and indexing would certainly be nice,
but seems far too expensive to contemplate for the foreseeable future
and will perhaps prove unnecessary.
It may well be possible to either automate the classification task or at
least provide authors assistance with this in the future, using
machine-learning techniques.
(See http://www.cora.justresearch.com for a search engine that also
classifies papers.)
for an example of one program that does this.)
In addition, I hope and expect that there will be
other overlay services that provide different organizations of the
material. It is easy to imagine services that define different
specialized sub­collections---for example, a collection tailored to
a specific university course or curriculum that possibly crosses
standard disciplinary boundaries and includes documents in a number of
repositories.
Critical to all such work are good mechanisms for
citations and cross-linking; this is currently an active research area.

I have mentioned a few facilities that could be added to CoRR
directly or provided by overlays. It is easy to imagine other
facilities that could be added or improved, such as an automatic
recommendation service. I suspect that some of the most
useful ones are the ones we cannot yet imagine. We are
optimistic that, given CoRR's open interface, these enhancements
and others will be investigated and implemented by the efforts of
many researchers in the community.  I expect that, as well, CoRR will at
some point grow to become, not just a repository for eprints, but for
other artifacts, such as data sets and tools.

Of course, CoRR is not the only repository in existence or being
contemplated. There are a number of efforts in other disciplines
to develop online scholarly publishing archives.  Significantly, there
has been a recent initiative to build an open architecture for online
repositories, so that all repositories can be linked together.  The
architecture will in fact be based on NCSTRL's Dienst protocol, just as
CoRR is.  This open architecture will permit the federation of online
repositories, and
could be an important step to creating an online Digital Library.

\subsection*{Appendix A: Committee Composition}

Ronald Boisvert, NIST \\
James Cohoon, Virginia\\
Peter Denning ({\em ex officio} -- former chair of the ACM Publications
Board)\\
Jon Doyle, MIT \\
Edward Fox, Virginia Tech\\
James Gray, Microsoft \\
Joseph Halpern, Cornell (chair)\\
Carl Lagoze, Cornell \\
Bernard Lang, INRIA\\
Michael Lesk, Bellcore \\
Steve Minton, ISI \\
Hermann Maurer, Graz, Austria \\
Andrew Odlyzko, ATT\\
Michael O'Donnell, U. Chicago\\
Bernard Rous, ACM \\
Jerome Saltzer, MIT \\
Erik Sandewall, Linkoping, Sweden\\
Stuart Shieber, Harvard \\
Jeffrey Ullman, Stanford \\
Rebecca Wesley, Stanford\\
Ian Witten, Waikato, New Zealand \\

\paragraph{Acknowledgments:}  I have already mentioned that this
article is based on two earlier articles.  I would again like to thank
Carl Lagoze, my coauthor on the second article, for his contributions.
I would also like to acknowledge the contributions of
the CoRR committee and the ACM publications board, especially Bill Arms
and Peter Denning, for their role in setting up and supporting
CoRR and their feedback on previous versions this article.
\end{document}